\documentclass[pre,twocolumn]{revtex4-1}

\usepackage{graphicx}
\usepackage{dcolumn}
\usepackage{bm}

\usepackage[utf8]{inputenc}
\usepackage[T1]{fontenc}
\usepackage{mathptmx}
\usepackage{bm}
\usepackage{pdfrender}

\usepackage{amsmath}
\usepackage{mathtools}
\usepackage{amssymb}

\usepackage{pst-node}
\usepackage{verbatim}   
\usepackage{color}      
\usepackage{subfigure}  
\usepackage{hyperref}   

\usepackage{braket} 

\newcommand{\bnabla}{\mbox{\boldmath $\nabla$}}

\newcommand{\ba}{\begin{eqnarray}}
\newcommand{\ea}{\end{eqnarray}}
\newcommand{\be}{\begin{equation}}
\newcommand{\ee}{\end{equation}}

\begin{document}

\title{Positing the problem of stationary distributions of active particles as third-order differential equation} 

\author{Derek Frydel}
\affiliation{Department of Chemistry, Universidad Técnica Federico Santa María, Campus San Joaquin, Santiago, Chile}

\date{\today}

\begin{abstract} 
In this work we obtain third order linear differential equation for stationary distributions of run-and-tumble particles 
in two-dimensions in a harmonic trap.  The equation represents the condition $j=0$ where $j$ is a flux and is obtained
from inference, using different known results in the limiting conditions.  Since analogous equation for passive Brownian 
particles is first order, a second and third order term must be a feature of active motion.  
In addition to formulating the problem as third order equation, we obtain solutions in the form of convolution of two 
distributions, Gaussian distribution due to thermal fluctuations, and the beta distribution
due to active motion at zero temperature.  The convolution form of the 
solution indicates that the two random processes are independent and the total distribution is the sum of those 
two processes.  

\end{abstract}

\pacs{
}

\maketitle

\section{Introduction}

Even for the case of an ideal-gas of active particles, exact expressions of stationary distributions remain sparse.  
One way that could potentially advance our understanding of this problem is to try to represent stationary 
distributions through a differential equation.  The structure of such an equation can shed light on the problem 
and point toward a possible resolution.  It could also play a classifying role and  be used to relate active particles 
to other physical systems or, in a more mathematical sense, to other systems of differential equations.  

In this work we obtain a third order linear differential equation for stationary distributions of run-and-tumble 
particles in two-dimensions in a harmonic trap 
\cite{Brady16,Cates08,Cates09,Martens12,Gomper15,Wagner17,Lowen17,Dhar19,Angelani19,Demery19,Basu20, Dhar20, 
Pruessner21, Pruessner22, Frydel22a}.    
No exact distributions are known for this system.  The terms of an equation represent flux $j$ and the equation 
itself represents the condition $j=0$.  In comparison, an equivalent equation for passive Brownian particles is 
of first order, where the first order term comes from diffusion.  The second and third order term, therefore, are 
features of active motion.

Third order differential equations are less commonly encountered in physics than second order ones \cite{Animalu10}, 
and so are the methods to solve them \cite{Hoeij07,Koepf19}.  This property can, at least partially, explain some of the 
difficulties in obtaining exact solutions.

Active particles are formulated and fully described by the Fokker-Planck equation, the solution of which
is the distribution $\rho$ in a phase-space.  For two-dimensions and in a steady-state, this distribution is a function 
of the particle position and the orientation of a swimming direction, $\rho\equiv \rho({\bf r},\theta)$.  
In this work, however, we are interested in the differential equation for the distribution $p$ as a function 
of a position only, $p\equiv p({\bf r})$, related to the distribution $\rho$ as $p = \int d\theta\,\rho$.  There is no 
straightforward way of transforming the Fokker-Planck equation into a differential equation for $p$.
Consequently, a considerable part of this work is devoted to the development of a method for obtaining 
such an equation.  


Self-propelled particles in a harmonic trap have been the subject of study in numerous papers.  
To highlight some more recent advancements, 
in \cite{Dhar20} the authors obtained exact series solution for a stationary distribution of active Brownian particles 
(ABP) in a harmonic trap in two-dimensions.  More recently, an expression for the entropy production 
rate has been obtained for the run-and-tumble model (RTP) in \cite{Pruessner21,Pruessner22,Frydel22a}. 
Dynamics of self-propelled particles in harmonic confinement were considered in \cite{Gomper15,Demery19}.

This paper is organized as follows.  In (\ref{sec:sec1}) we consider the RTP model in 1D.  
Given that many results in 1D are easier to obtain, the results of this section play the role of 
reference point for systems in higher dimensions.  
In (\ref{sec:sec2a}) we consider the RTP model in two-dimension in a harmonic potential 
in the direction along $x$-axis, $u = Kx^2/2$.  Differential equation for $p$ is obtained
together with the solution in form of convolution of two distributions.  
In (\ref{sec:sec2b}) we then consider a harmonic trap 
with isotropic symmetry, for which a differential equation for $p$ is also obtained together with 
a solution.  
In (\ref{sec:sec4}) we consider a harmonic trap in three-dimensions.  In this dimension, a
simple differential equation for $p$ is no longer possible.  Some analysis is carried out
to understand why this is so.  
In (\ref{sec:sec5}) the results are consolidated and discussed.  The work is concluded with 
(\ref{sec:sec6}).

\section{Setting up the stage:  RTP model in 1D}
\label{sec:sec1}

In the RTP model, particles, in addition to undergoing diffusion, move with velocity of constant 
magnitude $v_0$ and evolving in time orientation.  New orientations are assigned at discrete time 
intervals drawn from exponential distribution whose average value $\tau$ represents the persistence 
time.  Because in one-dimension there are two directions, the Fokker-Planck equation (FP) can be 
represented as a system of two coupled differential equations 
 \cite{Schnitzer93,Cates08,Cates09,Angelani17,Dhar18,Razin20,Frydel21b,Frydel22b}
\ba
&& \dot p_+  =  -v_0 p_+'   +   D p_+''  -   \frac{1}{2} \frac{1}{\tau} (p_+  -  p_-), \nonumber\\
&& \dot p_-   =   ~~~ v_0 p _-'   +   D p_-''   -   \frac{1}{2} \frac{1}{\tau} (p_-  -  p_+),
\label{eq:FPT-free-2}
\ea
where $p_+(x,t)$ and $p_-(x,t)$ are the distribution for particles with forward and backward swimming 
direction, respectively.  The first two terms in each equation come from the flux $j_{\pm} = -D p_{\pm}' \pm v_0$, 
where $D$ is the diffusion constant.  The last term in each equation gives rise to active motion and 
represents conversion of particles with one drift direction into particles with another drift direction.  
The conversion occurs with the rate $1/\tau$.   

The two equations in (\ref{eq:FPT-free-2}) can be combined into a single differential equation for the total 
distribution $p = p_+ + p_-$.   The procedure outlined in Appendix (\ref{sec:app1}) leads to the following 
result:  
\be
\frac{\partial}{\partial t} \bigg[ p   -   2D \tau p'' \bigg]     +     \tau \frac{\partial ^2 p}{\partial t^2}   =    -j', 
\label{eq:FPT-free-1}
\ee
where the flux $j = j_+ + j_-$ on the right hand side is given by 
\be
j = -(D  +   \tau v_0^2) p'   +  \tau D^2 p'''.  
\label{eq:flux-free}
\ee

Compared with the equivalent equation for passive Brownian particles governed by 
$
\dot p =   D p'', 
$
the first striking difference is a more complex time dependence that involves second order time derivative, 
the signature of ballistic motion.  This term is proportional to the persistence time $\tau$.  Furthermore, the 
terms that are first-order in time involve $p$ and $p''$.  

Turning next to the expression of flux in Eq. (\ref{eq:flux-free}), we find that it is modified in two ways.  
The first change is enhanced diffusion constant $D_{eff} = D + \tau v_0^2$.  The second and more significant modification 
is the emergence of a third-order term which has no counterpart in passive Brownian motion 
and as such can be regarded as a signature of active motion.  
Note that in the limit $\tau\to 0$, when the drift direction changes extremely fast, active motion becomes
negligible and Eq. (\ref{eq:FPT-free-1}) recovers the standard diffusion equation.  

At a steady-state, the total flux vanishes, $j=0$, resulting in the following equation:  
\be
0 = -(D  +   \tau v_0^2) p'   +  \tau D^2 p'''.  
\label{eq:j-free}
\ee
A stationary distribution $p$ is then a solution to a third-order differential equation.  

For particles confined between two walls at $x= \pm h$, to solve Eq. (\ref{eq:j-free}) 
we need the following boundary conditions:  
\ba
&& 0 = p'(0),  \nonumber\\
&& 0  =  -v_0^2  p(\pm h) + D^2 p''(\pm h).  
\ea
The first condition ensures that $p$ is symmetric around $x=0$ and the second one that the fluxes 
$j_{\pm}=-D p' \pm v_0$ vanish separately at the walls.  (In contrast to $j$, the constituent fluxes 
$j_+$ and $j_-$ are not required to vanish everywhere in the interval.  This gives rise to internal currents 
responsible for non-zero entropy production rate \cite{Razin20,Frydel22a}).  The solution to Eq. (\ref{eq:j-free}) 
turns out to have a simple functional form: $p(x) = a + b\cosh(kx)$ \cite{Razin20,Frydel22b}.

\subsection{Harmonic potential}
\label{sec:sec1a}

For the case of a harmonic confinement, represented by the potential 
$u(x) = K x^2 / 2$, the two coupled FP equations for the RTP model in 1D are 
\ba
&& \dot p_+  =  
\left[ \left( \mu K x - v_0 \right) p_+\right]'    +  D p_+''  -  \frac{1}{2} \frac{1}{\tau} (p_+  -  p_-), \nonumber\\
&& \dot p_-  =  
 \left[ \left( \mu K x + v_0\ \right) p_- \right]'  +  D p_-''   -   \frac{1}{2} \frac{1}{\tau} (p_-  -  p_+). 
\label{eq:FP_RTP1Da}
\ea
To simplify expressions, we define the following time and length scales 
$
\tau_k = \frac{1}{\mu K}$, $\lambda_k = v_0 \tau_k,
$
and from now on work with dimensionless space and time variables $z = \frac{x}{\lambda_k}$, 
$s = \frac{t}{\tau_k}$.   The dimensionless diffusion constant and the rate of orientational change
 are defined as
$
B = \frac{D\tau_k }{\lambda_k^2}$ and $\alpha = \frac{\tau_k}{\tau}.  
$
The two equations in (\ref{eq:FP_RTP1Da}) become 
\ba
&& \dot p_+  =  
\left[ \left( z - 1 \right) p_+\right]'    +  B p_+''  -  \frac{\alpha}{2}  (p_+  -  p_-), \nonumber\\
&& \dot p_-  =  
 \left[ \left( z + 1\ \right) p_- \right]'  +  B p_-''   -   \frac{\alpha}{2} (p_-  -  p_+). 
\label{eq:FP_RTP1D}
\ea
Combining the two equations into a single differential equation, following the procedure 
in Appendix (\ref{sec:app2}), leads to  
\be
\frac{\partial }{\partial s} \bigg[ (\alpha-3) p    -  2z p'  -  2B p''\bigg]  +  \frac{\partial ^2 p}{\partial s^2}    =  - j',  
\label{eq:FPT-harmonic-1}
\ee
with the flux given by 
\be
j =    -(\alpha-2)zp    -   (1 - z^2 - 3B + B\alpha)p'    +    2B zp''    +   B^2 p'''. 
\label{eq:flux-harmonic}
\ee
Compared to passive particles, the stiffness parameter is renormalized as $K_{eff} = (\alpha-2) K$.  
What is interesting is that for $\alpha<2$ the stiffness becomes negative, causing particles to be 
repelled from the trap center rather than being attracted to it.  For those values of $\alpha$, the distribution
is bimodal with two symmetric peaks shifted away from the center.   If we interpret the coefficient of the first 
order term as $z$-dependent effective diffusion, $B_{eff} = (1 - z^2 - 3B + B\alpha)$, then there is $|z|$ 
beyond which $B_{eff}$ is negative.   Without the second and third order terms, the physical distribution 
does not exist beyond this point.  Including these higher order terms allows the distribution to extend 
beyond this point.  The peak centers, however, remain confined into the region where $B_{eff} >0$, 
based on what is seen from calculated $p$.  For $\alpha>2$, the distribution $p$ becomes unimodal with 
a single peak at $z=0$. 

At a steady-state, $j=0$ and a stationary distribution is obtained by solving 
\ba
0 =    (2-\alpha)zp    -   (1-z^2  - 3B  +  B\alpha)p'    +    2B zp''    +   B^2 p'''.  \nonumber\\
\label{eq:FP0-harmonic}
\ea

We are next going to consider the solution $p$ in two limiting situations:  without thermal fluctuations (with
active motion only), and without active motion (with thermal fluctuations only).  At zero temperature 
($B=0$) and without thermal fluctuations, Eq. (\ref{eq:FP0-harmonic}) reduces to a first-order differential equation:  
\be
0 = (2-\alpha)zp    -   (1-z^2)p', 
\label{eq:p00_1D}
\ee
for which the solution is a beta distribution on the interval $[-1,1]$ \cite{Cates08,Cates09,Dhar19,Basu20}:  
\be
p_b \propto (1-z^2)^{\frac{\alpha}{2}-1},
\label{eq:p00_1D-sol}
\ee
where we use the subscript $b$ to indicate that $p$ is represented by a beta distribution.
An active motion becomes suppressed in the limit $\alpha\to \infty$ as a result of rapid alternation 
of a swimming direction.   Eq. (\ref{eq:FP0-harmonic}) in this situation reduces to 
\be
0 =    -zp    -    B p',
\label{eq:p_1D-alpha}
\ee
and the solution is given by the Boltzmann distribution 
\be
p_g\propto e^{-z^2/2B},
\ee
where we use the subscript $g$ to indicate that $p$ is represented by a Gaussian distribution.

There is no closed form solution to Eq. (\ref{eq:FP0-harmonic}) we are aware of, but it turns out that 
the solution can be expressed as convolution of two limiting probability distributions discussed above, 
the beta distribution in Eq. (\ref{eq:p00_1D-sol}) and the Boltzmann distribution for the limit \cite{JML22}:  
\be
p(z) \propto \int_{-1}^{1} dz'\, (1-z'^2)^{\frac{\alpha}{2}-1} e^{-\frac{(z-z')^2}{2B}}.
\label{eq:p1D-conv}
\ee
See Appendix (\ref{sec:app0}) for derivation.

Before dwelling into physical interpretation of the mathematical form of the solution in Eq. (\ref{eq:p1D-conv}), 
we briefly consider some mathematical aspects of it.  
Because in the limit $B\to 0$ the Gaussian function becomes a delta function, 
$
\lim_{B\to 0} e^{-{(z-z')^2 } / { 2B}} \propto \delta(z-z'), 
$
the solution in Eq. (\ref{eq:p1D-conv}) recovers the solution in Eq. (\ref{eq:p00_1D-sol}).  
And because the beta distribution defined on the interval $[-1,1]$ in the limit $\alpha\to\infty$ 
also becomes a delta function, 
$
\lim_{\alpha\to \infty} (1-z'^2)^{\frac{\alpha}{2}-1} \propto  \delta(z'), 
$
the solution in Eq. (\ref{eq:p1D-conv}) correctly recovers the Boltzmann distribution.

Next we are going to reflect on physical interpretation of the solution in Eq. (\ref{eq:p1D-conv}).  
The convolution of two distributions arises for the process that involves a sum of two 
or more independent random variables.  For example, if the random variables $x$ 
(with the associated distribution $p_x(x)$) and $y$ (with the associated distribution $p_y(y)$)
are independent, then the distribution for the process $z=x + y$ is $p_z(z) = \int dz' \, p_x(z')p_y(z'-z)
=  \int dz' \, p_y(z')p_x(z'-z)$.  In light of this, the solution in Eq. (\ref{eq:p1D-conv}), which 
can be represented as $p = \int dz' \, p_b(z') p_g(z'-z)$, makes 
perfect sense, since the two random processes, thermal fluctuations and active motion, 
are independent, and the distribution $p$ that we are looking for is for the sum of those
two processes.  One consequence 

This behavior, however, should not be taken as general, valid for any external potential.  
It is rather a unique feature of harmonic potentials, or linear external forces.   Despite this, in the 
context of harmonic potentials, this is an extremely useful insight.  


\subsection{Other exact results}

In this section, we briefly consider other exact results for the system at hand.  These results may 
appear superfluous at this stage but will make more sense once we consider  RTP particles in two-dimensions.  
The reader may skip this section and return to it later.  

One of the quantities that we are going to find useful in analyzing harmonic trap in higher dimensions are 
the even moments $\langle z^{2n}\rangle=\int_{-1}^{1} dz\, z^{2n} p(z)$ for the zero temperature limit.  
For the system in 1D we know that the distribution in this limit corresponds to beta distribution in Eq. (\ref{eq:p00_1D-sol}), 
and the moments can be calculated as  
\be
\langle z^{2n}\rangle = \prod_{m=1}^n \frac{2m-1}{2m-1 + \alpha}, ~~~~ n = 1,2,\dots.  
\label{eq:zn2}
\ee
Alternatively the moments can be calculated directly from Eq. (\ref{eq:p00_1D}) 
by operating on it with the integral operator $\int_{-1}^{1} dz\, z^{2n-1}$.  This leads to the following iterative expression 
\be
\langle z^{2n}\rangle  = \frac{2n-1}{2n - 1 + \alpha} \langle z^{2n-2}\rangle.  
\label{eq:zn2b}
\ee

Another limiting situation of interest is the limiting case $\alpha=0$. Eq. (\ref{eq:FP0-harmonic}) 
in this case reduces to 
\be
0 =    2zp    -   (1-z^2  - 3B)p'    +    2B zp''    +   B^2 p'''.  
\label{eq:FP0-harmonic-q}
\ee
The solution for this case can be inferred from physical 
considerations.  Since for $\alpha=0$ swimming orientations do not evolve in time, each particle with its fixed 
orientation attains equilibrium.  
But because there are two swimming orientations, the system is a mixture of particles with different 
swimming orientations --- the case of quenched disorder \cite{Frydel21a,Frydel22a}. Accordingly, a distribution 
is represented as a superposition of Boltzmann distributions of particles with different swimming orientations, 
in dimensionless units given by 
$
p(z) \propto e^{ -\frac{z^2}{2B} } e^{ \frac{z}{B} } + e^{ -\frac{z^2}{2B} } e^{ -\frac{z}{B} }.  
$

\section{ RTP model in 2D}
\label{sec:sec2} 

For RTP particles in a harmonic trap in two or higher dimensions, the task of reducing the corresponding stationary 
FP equation into a single differential equation for the distribution $p$ is more challenging.  In 1D, things 
are simplified on account of there being only two discrete orientations.  For higher dimensions, there are infinitely many 
orientations.

\subsection{Harmonic potential $u = Kx^2/2$}
\label{sec:sec2a} 

In this section we consider the RTP particles in 2D trapped in a harmonic potential of the form $u(x) = Kx^2/2$.  The system 
is effectively one-dimensional as it is translationally invariant along the $y$-axis.  The swimming orientation is 
specified by the unit vector ${\bf n} = (\cos\theta,\sin\theta)$ and the relevant distribution is $\rho(x, \theta, t)$
normalized as $\int_{-\infty}^{\infty} dx\int_0^{2\pi} d\theta\, \rho = 1$.

The stationary Fokker-Planck equation that determines the distribution $\rho(x, \theta)$ in dimensionless units is given by
\be
0  =  \left[ \left( z - \cos\theta \right) \rho\right]'    +  B \rho''  -  \alpha \left( \rho - \int_0^{2\pi} \frac{d\theta}{2\pi}\, \rho \right). 
\label{eq:FP0-2DL}
\ee
See Appendix (\ref{sec:app4}) for details. 


The procedure in Appendix (\ref{sec:app2}) is not generalizable to continuous orientations.  
Consequently, another procedure is required.  
We begin by considering the zero temperature limit, or $B=0$.  Eq. (\ref{eq:FP0-2DL}) in this case reduces to 
\be
0 =   z \rho'   +  \rho  -  \cos\theta \rho'  -  \alpha \rho    +    \frac{\alpha }{2\pi} p.  
\label{eq:diff-1}
\ee
where $p = \int d\theta\,\rho$.  We next use this equation to generate expressions for even moments 
$\langle z^{2n}\rangle = \int_{-1}^1 dz\int_0^{2\pi} d\theta \, z^{2n} \rho$, in analogy to Eq. (\ref{eq:zn2b}).   
For example, to obtain an expression for $\langle z^{2}\rangle$, we 
operate on Eq. (\ref{eq:diff-1}) with the integral operator $\int_{-1}^1 dz\int_0^{2\pi} d\theta \, z^2$.  This 
yields 
$
\langle z^{2}\rangle = \langle z\cos\theta\rangle. 
$ 
Then to obtain $\langle z\cos\theta\rangle$, we operate on Eq. (\ref{eq:diff-1}) with 
$\int_{-1}^1 dz\int_0^{2\pi} d\theta \, z\cos\theta$, which yields $\langle z\cos\theta\rangle = 1/2(1+\alpha)$, 
so that we can write 
$$
\langle z^2 \rangle = \frac{1}{2} \frac{1}{1+\alpha}.  
$$
A similar procedure can be used obtain $\langle z^4\rangle$ and any other moment.  From the sequence of
such moments, one can then infer a general formula 
\be
\langle z^{2n} \rangle = \prod_{k=1}^n \frac{1}{2} \frac{1}{n+\alpha}, ~~~~ n = 1,2,\dots.  
\label{eq:zn2c}
\ee
that alternatively can be represented as an iterative relation given by 
\be
\langle z^{2n} \rangle =  \frac{1}{2} \frac{2n-1}{n+\alpha}  \langle z^{2n-2} \rangle.  
\label{eq:zn2d}
\ee

We attempt next to infer a differential equation that generates such moments.  Similarity between the result 
in Eq. (\ref{eq:zn2d})and that in Eq. (\ref{eq:zn2b}) suggests that a differential equation ought to have a similar 
structure to Eq. (\ref{eq:p00_1D}).  We find that the equation 
\be
0 = (1 - 2\alpha)zp    -   (1-z^2)p',
\label{eq:p00_2DL}
\ee
generates the moments in Eq. (\ref{eq:zn2d}).  
The solution is a beta distribution on the interval $[-1,1]$, similar to that in Eq. (\ref{eq:p00_1D-sol}) but
that scales differently with $\alpha$, 
\be
p(z) \propto (1-z^2)^{\alpha-\frac{1}{2}}.  
\label{eq:p0_2DL}
\ee

We next consider the case $\alpha=0$, the limiting case where swimming orientations no longer
evolve in time and because particles are confined, the system attains equilibrium.  Yet because 
the system is a mixture of particles with different swimming orientations, it represents 
quenched disorder where a stationary distribution is a superposition of Boltzmann distributions of 
particles with different swimming orientations.  

For a given orientation $\theta$, a normalized Boltzmann distribution is 
\be
\rho(x,\theta) = \left[ \sqrt{\frac{\beta K}{2\pi}}  e^{-\frac{v_0^2 \cos^2\theta}{2\beta KD^2}} \right]  
e^{ - \frac{\beta Kx^2}{2} }  e^{  \frac{v_0 x \cos\theta }{D} }. 
\label{eq:rho-alpha}
\ee
Integrating the above expression over all orientations yields the distribution $p$, 
which, in dimensionless units becomes 
\be
p(z) \propto  e^{ - \frac{z^2}{2B} }  \int_{0}^{2\pi} d\theta\,   e^{-\frac{ \cos^2\theta}{2 B}}  e^{  \frac{z \cos\theta }{B} }.  
\label{eq:p2DL-q}
\ee
We are now in a situation where we know the solution but do not have a differential equation that 
generates it.  This constitutes an inverse problem to that for determining a solution from a known equation.  
We find the following differential equation generates the solution of interest 
\be
0 = z p   -   (1 - z^2 - 2B) p'    +   2 B z p''   +   B^2 p'''.  
\label{eq:diff-2DL-q} 
\ee
The procedure used to infer this equation is described in Appendix (\ref{sec:app3}).

At this point we have two equations for two limiting situations, Eq. (\ref{eq:p00_2DL}) for $B=0$ 
and Eq. (\ref{eq:diff-2DL-q}) for $\alpha=0$.   We combine the two equations in such a way as to 
avoid repetition of the same terms: $zp$ and $(z^2-1)p'$.  The resulting equation still fails to 
recover $0 = -zp - Bp'$ in the limit $\alpha\to\infty$.  
This can be fixed by including an additional term, $-2\alpha B p'$.  Note that this is the only 
term that couples active and diffusive motion.  The complete third order  equation for stationary 
$p$ is 
\be
0 = (1 - 2\alpha) z p   -   (1 - z^2 - 2B + 2\alpha B) p'    +   2 B z p''   +   B^2 p'''.  
\label{eq:diff-2DL}
\ee
Note that Eq. (\ref{eq:diff-2DL}) has the same structure as Eq. (\ref{eq:FP0-harmonic}) for the true 
1D system.  In fact, the two equations can be scaled into each other if for every $\alpha$ in Eq. (\ref{eq:diff-2DL}) 
we substitute $(\alpha-1)/2$.  
This  implies that the solution in Eq. (\ref{eq:p1D-conv}) can be scaled to produce a solution to 
Eq. (\ref{eq:diff-2DL}): 
\be
p_{}\left(z;\alpha\right) = p_{1D}\left(z;2\alpha+1\right),
\ee
or more specifically
\be
p(z) \propto \int_{-1}^{1} dz'\, (1-z'^2)^{\alpha-\frac{1}{2}} e^{-\frac{(z-z')^2}{2B}}.
\label{eq:p2DL-conv}
\ee
The differential equation in Eq. (\ref{eq:diff-2DL}) and the solution in Eq. (\ref{eq:p2DL-conv}) are the 
main results of this section.

\begin{figure}[hhhh] 
 \begin{center}
 \begin{tabular}{rrrr}
\includegraphics[height=0.20\textwidth,width=0.24\textwidth]{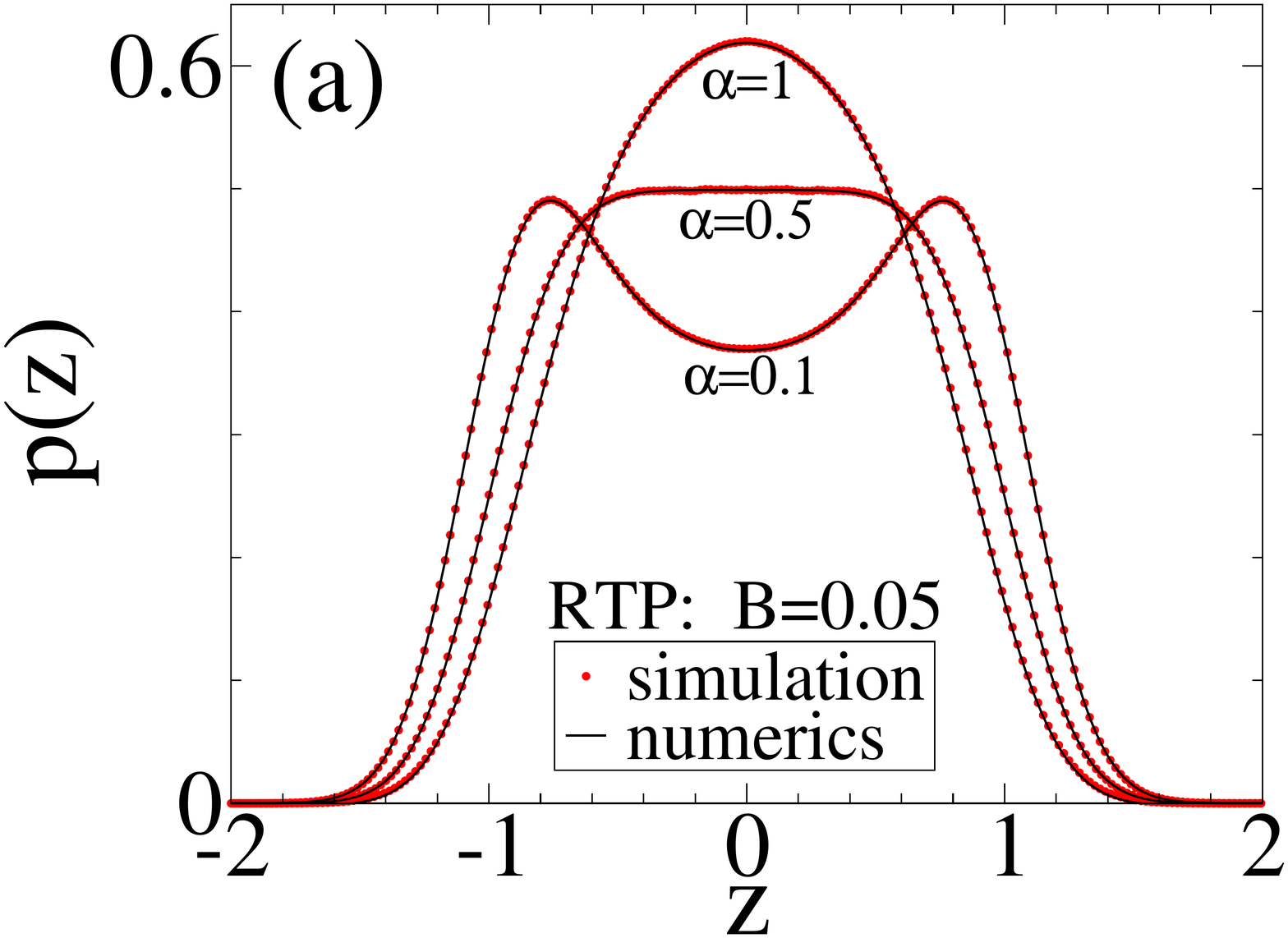}  
\includegraphics[height=0.20\textwidth,width=0.24\textwidth]{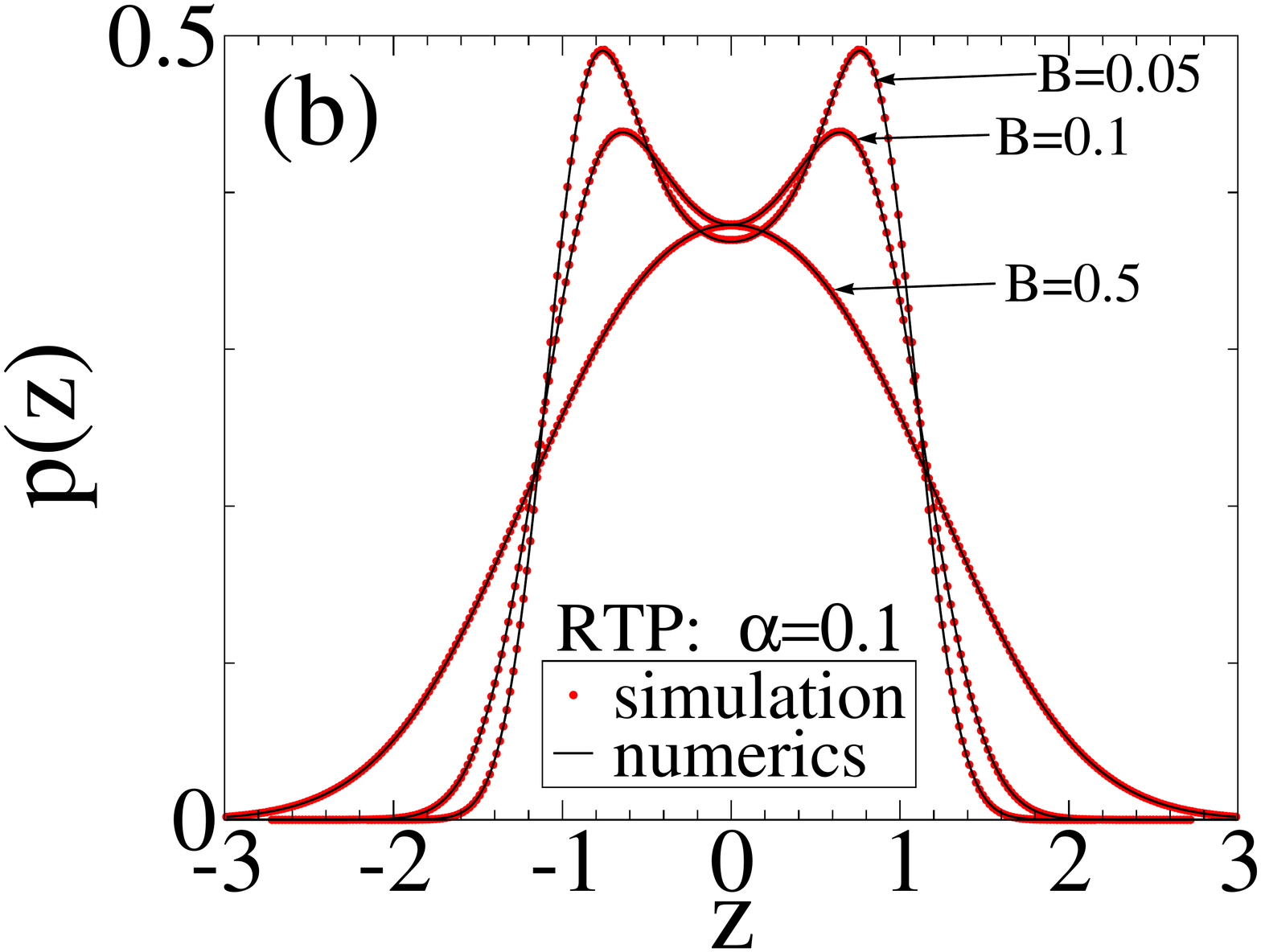}  
 \end{tabular}
 \end{center} 
\caption{Stationary distributions for RTP particles in 2D in a harmonic trap $u = Kx^2/2$.  
Solid lines correspond to the solution in Eq. (\ref{eq:p2DL-conv}) evaluated numerically.
Points represent simulation data.  Simulations were carried using the Euler 
method for updating particle positions:  $z(t+\Delta t) = \left[ \cos\theta(t) - z(t)\right]\Delta t + \hat\zeta(t) \sqrt{2B\Delta t}$,
where $\hat\zeta$ is a Gaussian noise with zero mean and unit variance.
}
\label{fig:fig1} 
\end{figure}

To verify Eq. (\ref{eq:diff-2DL}), 
we compare $p$ obtained from numerical evaluation of Eq. (\ref{eq:p2DL-conv}) with stationary 
distributions obtained from simulations.  The results are plotted and compared in Fig. (\ref{fig:fig1}).  
The agreement between the two methods provides confirmation for the correctness of Eq. (\ref{eq:diff-2DL}).
\graphicspath{{figures/}}

\begin{figure}[hhhh] 
 \begin{center}
 \begin{tabular}{rrrr}
\includegraphics[height=0.18\textwidth,width=0.23\textwidth]{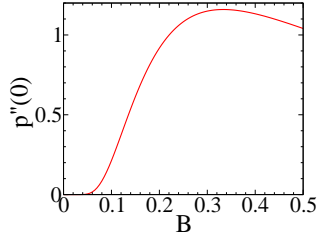}  
 \end{tabular}
 \end{center} 
\caption{ Second derivative of the distribution $p$ at $z=0$ and calculated for $\alpha_c=1/2$, 
see Eq. (\ref{eq:ddp}), as a way to quantify the degree of flatness.  }
\label{fig:fig1b} 
\end{figure}

Note that the differential equation in Eq. (\ref{eq:diff-2DL}) and the corresponding solution in Eq. (\ref{eq:p2DL-conv}) 
simplify at a specific value of $\alpha_c=1/2$.    At this crossover value of $\alpha$ the effective stiffness of a trap 
$K_{eff}=(2\alpha-1)K$ changes sign.  As a consequence, the zero order term in Eq. (\ref{eq:diff-2DL}) vanishes and 
the integral solution in Eq. (\ref{eq:p2DL-conv}) simplifies to yield 
\be
p(z) \propto  \text{erf} \left[\frac{z+1}{\sqrt{2 B}}\right] -  \text{erf} \left[\frac{z-1}{\sqrt{2 B}}\right].  
\label{eq:ddp}
\ee
At $B=0$, the above distribution becomes a rectangular function.  Then as $B$ increases, 
it transforms into a normal distribution.   A degree of flatness of this distribution can be quantified from the 
second derivative at $z=0$, which is found to be $p''(0) \propto 1/\sqrt{B^3 e^{1/B}}$ and that is plotted 
in Fig. (\ref{fig:fig1b}).  The distribution can be assumed to be flat if $p''(0)\approx 0$.  This is the case for 
$B \lesssim 0.06$, then for larger $B$, $p''(0)$ exhibits sharp increase.  
\graphicspath{{figures/}}


Fig. (\ref{fig:fig1}) (a) shows distributions before, at, and beyond the crossover $\alpha_c$.  All the plots are 
for $B=0.05$, and so the distribution is still considered as flat at the crossover.  
Fig. (\ref{fig:fig1}) (b) shows distributions at a fixed $\alpha$, below the crossover value, for different values of $B$.  
For $B=0.5$ thermal fluctuations completely destroy the bimodal structure and the distribution  
becomes unimodal.

\subsection{ Harmonic potential $u = Kr^2/2$}
\label{sec:sec2b} 

We next consider RTP particles (in 2D) in a harmonic trap with circular symmetry, $u(r) = Kr^2/2$.  
The stationary Fokker-Planck equation that determines the distribution $ \rho(z,\theta)$ in this case is 
\ba
0  &=&      z\rho' + 2 \rho -  \cos\theta \rho'  +  \frac{\sin\theta}{z} \frac{\partial \rho}{\partial \theta}   
\nonumber\\
&+& B\left[ \rho''       +   \frac{\rho'}{z}     +   \frac{1}{z^2} \frac{\partial^2 \rho}{\partial\theta^2}   \right]
  -  \alpha \left( \rho - \int_0^{2\pi} \frac{d\theta}{2\pi}\, \rho \right), 
\label{eq:FPS-2Da}
\ea
where $z = r/\lambda_k$.  In this scenario $\theta$ represents a relative angle between the swimming 
orientation and the angular position of a particle in a trap.  See Appendix (\ref{sec:app4}) for details of 
the derivation.

At zero temperature, or $B=0$, the stationary distribution $\rho$ is governed by 
\ba
0 =    z\rho' + 2 \rho 
-  \cos\theta \rho'  +  \frac{\sin\theta}{z} \frac{\partial \rho}{\partial \theta}   -  \alpha \rho +  \frac{\alpha}{2\pi} p.  
\label{eq:p-2DS-B}
\ea
where $p = \int_0^{2\pi} d\theta\,\rho$.  By operating on the above equation with integral operators 
of the form $2\pi \int_0^{2\pi} d\theta \int_0^1 dz \, z^{2n+1} \rho$ 
we can calculate the moments $\langle z^{2n}\rangle$.  
For example, to obtain $\langle z^2 \rangle = 2\pi \int_0^{2\pi} d\theta \int_0^1 dz \, z^{3} \rho$, we operate on 
Eq. (\ref{eq:p-2DS-B}) with $2\pi \int_0^{2\pi} d\theta \int_0^1 dz \, z^{3}$.   This yields 
$\langle z^2 \rangle   =   \langle z\cos\theta\rangle$.  Then to calculate $ \langle z\cos\theta\rangle$, 
we operate on Eq. (\ref{eq:p-2DS-B}) with $2\pi \int_0^{2\pi} d\theta \int_0^1 dz \, z^{2}\cos\theta$.  This leads to 
$$
\langle z^2 \rangle = \frac{1}{1+\alpha}.  
$$
A similar procedure can be used to calculate higher moments from which 
it is possible to infer the following general expression 
\be
\langle z^{2n}\rangle = \prod_{m=1}^n \frac{m}{m + \alpha},
\ee
alternatively expressed as an iterative relation 
\be
\langle z^{2n}\rangle  = \frac{n}{n + \alpha} \langle z^{2n-2}\rangle. 
\ee
The first-order differential equation that generates such moments is 
\be
0 = (2 - 2\alpha) z p + \left( z^2 -1 \right) p'. 
\label{eq:p00_2D-sph}
\ee
The solution of the above equation 
is a beta distribution on the interval $[-1,1]$, similar to that in Eq. (\ref{eq:p00_1D-sol}) 
and Eq. (\ref{eq:p0_2DL}) but
that scales differently with $\alpha$, 
\be
p(z) \propto (1-z^2)^{\alpha-1}.  
\label{eq:p0_2D}
\ee

We next consider the limiting case $\alpha=0$.  Because swimming orientations in this limit do not 
change in time, the confined RTP particles attain equilibrium, and because the system is a mixture
of particles with different swimming orientations, it represents quenched disorder
where a stationary distribution is a superposition of different Boltzmann distributions.  
For particles with swimming orientation $\theta$, the normalized Boltzmann distribution is that same as that 
in Eq. (\ref{eq:rho-alpha}) with $x\to r$.  
Integrating this distribution over all orientations, in dimensionless units yields 
\be
p(z) \propto e^{-\frac{z^2}{2B}} \,\,  {\text I}_0\left( \frac{z}{B} \right).  
\label{eq:p0}
\ee
The solution is the product of a Gaussian and modified Bessel function of the 
first kind.  The third-order differential equation that generates this solution is determined to be 
\be
0 =   2 z p  + \bigg( z^2 -1 +  4B  -  \frac{B^2}{z^2}\bigg) p'  +   \bigg(\frac{B^2}{z^2} + 2B \bigg) z p''  +  B^2 p'''.
\label{eq:p_2DS-B}
\ee
See Appendix (\ref{sec:app3}) for details.  

To obtain the complete equation, Eq. (\ref{eq:p00_2D-sph}) and Eq. (\ref{eq:p_2DS-B}) are combined 
taking care that the similar terms are not repeated.  To ensure the correct behavior in the limit $\alpha\to\infty$
we add the term $-2\alpha B p'$.  The complete equation becomes 
\ba
0 &=&   2(1-\alpha) z p   -  \bigg( 1 - z^2  -  4B  + 2\alpha B  +  \frac{B^2}{z^2} \bigg) p'  \nonumber\\
&& +   \bigg(2 B + \frac{B^2}{z^2}  \bigg) z p''  +  B^2 p'''.
\label{eq:p_2DS}
\ea

Using previously gained insights, it is expected that the solution to the above equation can be 
be represented as a convolution of the beta solution in Eq. (\ref{eq:p0_2D}) and the Gaussian 
distribution.  If we consider that the convolution is done in 2D space, this leads to the following 
results 
\be
p \propto  \int_0^1  dz'\, (1-z'^2)^{\alpha-1} e^{-\frac{ (z' - z)^2}{2B}}  \left[ z'  e^{-zz'/B} {\text I}_0\left(\frac{z z'}{B}\right) \right] .  
\label{eq:p2D-conv}
\ee
where ${\text I}_0(x)$ is the modified Bessel function of the first kind.  See Appendix (\ref{sec:app0}) for details.  

\begin{figure}[hhhh] 
 \begin{center}
 \begin{tabular}{rrrr}
\includegraphics[height=0.20\textwidth,width=0.24\textwidth]{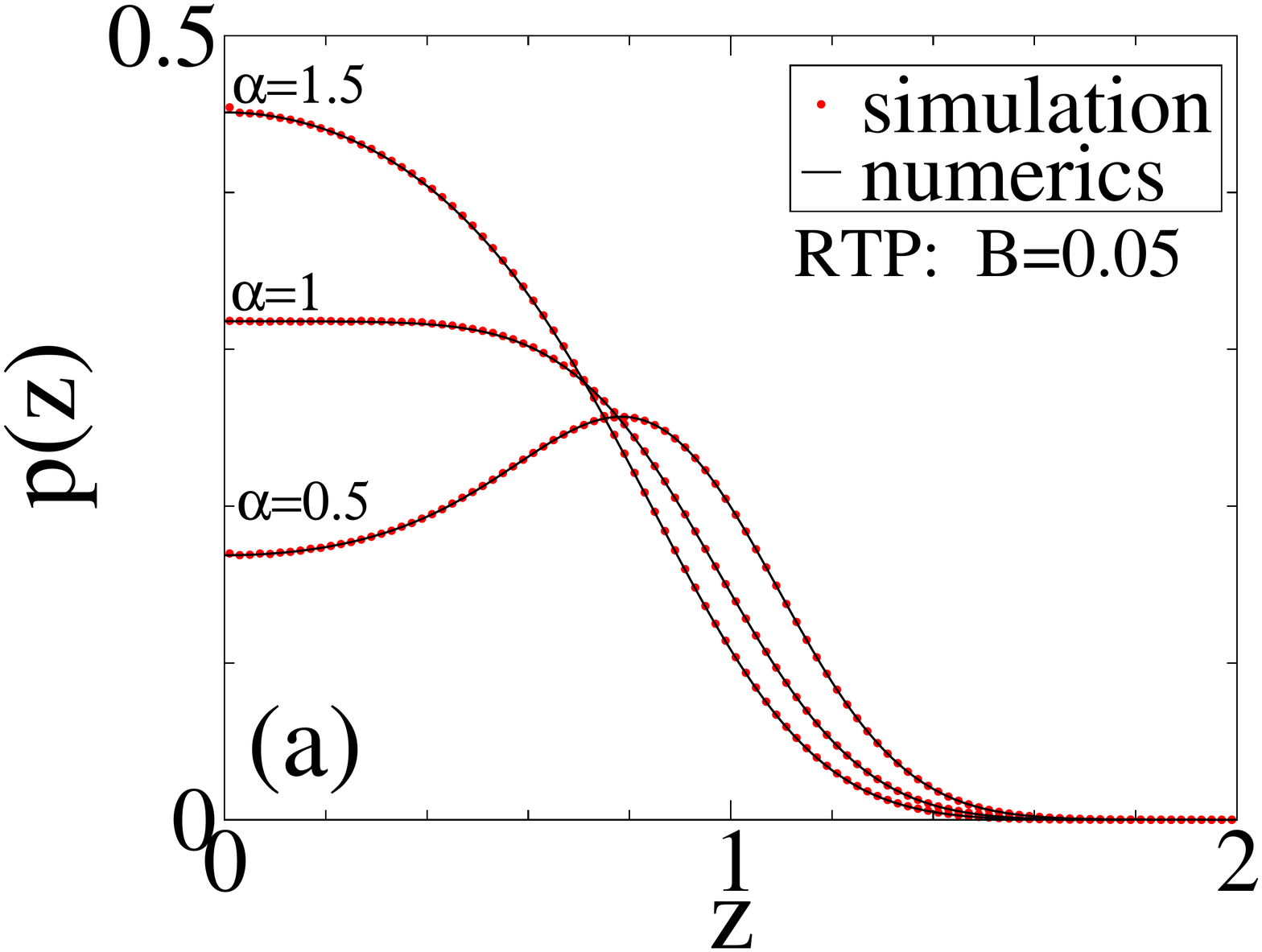}  
\includegraphics[height=0.20\textwidth,width=0.24\textwidth]{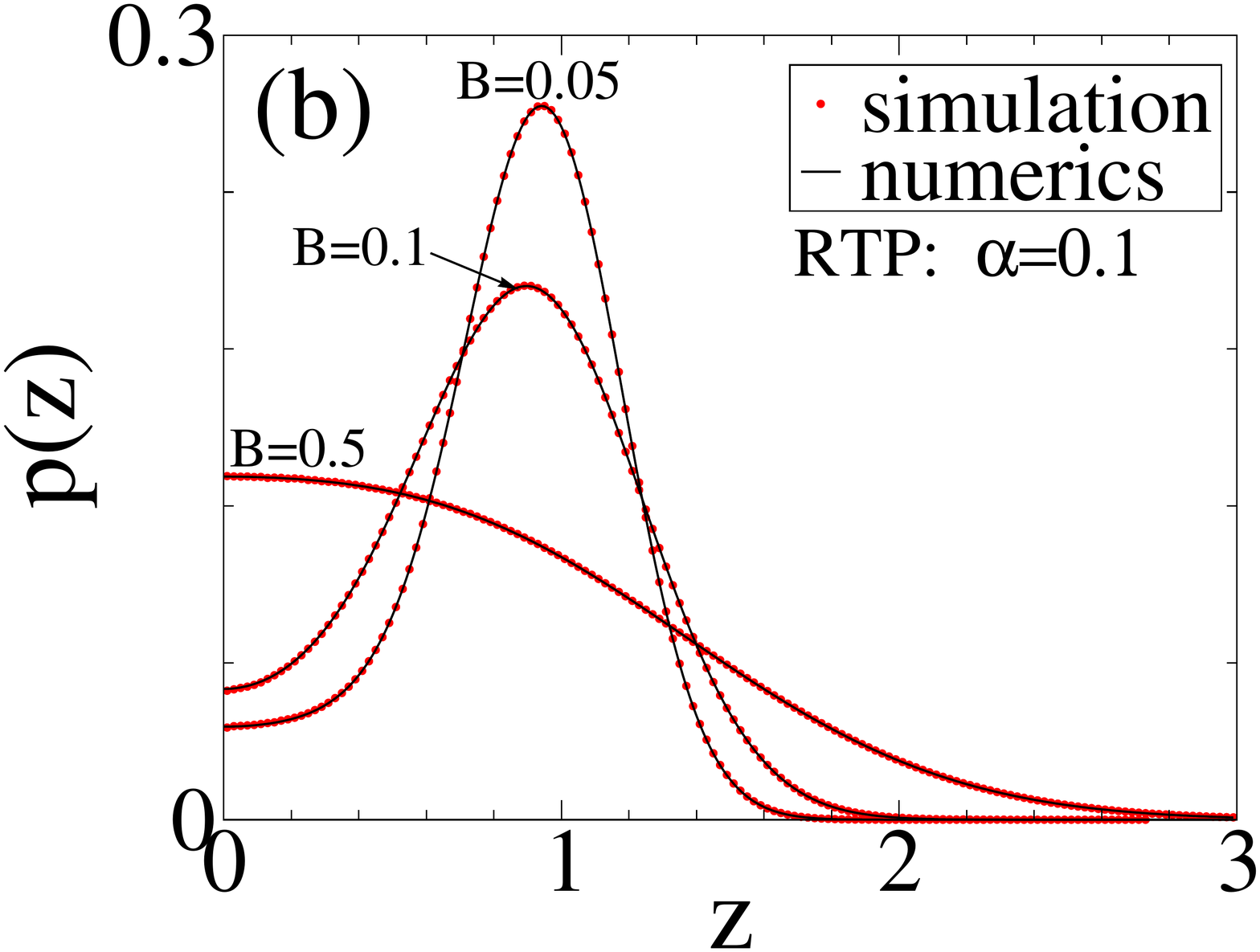}  
 \end{tabular}
 \end{center} 
\caption{Stationary distributions of RTP particles in 2D in an isotropic harmonic trap $u = Kr^2/2$.  
Solid lines correspond to numerical solution of Eq. (\ref{eq:p2D-conv}), 
and the points represent simulation data. 
}
\label{fig:fig2} 
\end{figure}

Eq. (\ref{eq:p_2DS}) together with the solution in Eq. (\ref{eq:p2D-conv})
are the main results of this section. 
To verify Eq. (\ref{eq:p_2DS}), in Fig. (\ref{fig:fig2}) we compare $p$ obtained from evaluating Eq. (\ref{eq:p2D-conv})
with the simulated data points.  The agreement is exact apart from statistical noise.  
\graphicspath{{figures/}}

\begin{figure}[hhhh] 
 \begin{center}
 \begin{tabular}{rrrr}
\includegraphics[height=0.18\textwidth,width=0.23\textwidth]{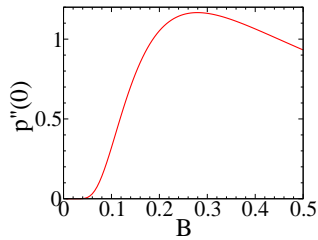}  
 \end{tabular}
 \end{center} 
\caption{ Second derivative of the distribution $p$ at $z=0$ and for $\alpha_c=1$, 
as a way to quantify the degree of flatness.  }
\label{fig:fig2b} 
\end{figure}

At the crossover value of $\alpha,$ $\alpha_c=1$, the zero order term in Eq. (\ref{eq:p_2DS}) vanishes and the
solution becomes 
\be
p(z) \propto  e^{-\frac{z^2}{2B}}  \int_0^1  dz'\, z' e^{-\frac{ z'^2}{2B}} {\text I}_0\left(\frac{z z'}{B}\right).  
\label{eq:ddpc}
\ee
The distribution becomes rectangular for $B=0$, then with the onset of thermal fluctuations it approaches a
normal distribution.  Fig. (\ref{fig:fig2b}) shows how the second derivative of this distribution measured at $z=0$ 
changes with $B$.    
\graphicspath{{figures/}}

Fig. (\ref{fig:fig2}) (a) plots the distribution at $\alpha_c=1$ and two other 
distributions on both sides of the crossover.  All distributions are for $B=0.05$ when thermal 
fluctuations are still small.   To demonstrate the effect of temperature, Fig. (\ref{fig:fig2}) (b) shows the 
distributions for a fixed $\alpha$, below the crossover value, for three different $B$.

%
%
%
%
%
%

\section{ RTP model in 3D:  considerations}
\label{sec:sec4} 

It turns out that in the case of the RTP model in 3D it is no longer possible to obtain a simple differential equation for 
the stationary distribution $p$, at least within the methodology developed and used in previous sections.  The main 
obstacle seems to be the fact that it is not obvious how to infer a differential equation for the zero temperature limit, 
or $B=0$, from the moments.  

To illustrate those difficulties, we consider a harmonic potential along the $x$-axis, $u(x)=Kx^2/2$.  The stationary FP 
equation is the same as that in (\ref{eq:FP0-2DL}).  What is different is the way how the averaging over a swimming 
orientation is evaluated:  
\be
p(z) = \int_0^{\pi} d\theta\, \sin\theta \,  \rho(z,\theta).  
\ee
This apparently trivial modification, in actuality turns out to complicate things.  Recall that for the system in 2D and 
for the same potential the averaging is done as $p(z) = \int_0^{2\pi} d\theta\,  \rho(z,\theta)$.

At zero temperature or $B=0$, the stationary FP equation becomes:  
$$
0 =   z \rho'   +  \rho  -  \cos\theta \rho'   -  \alpha \rho    +    \frac{\alpha }{2} p.  
$$
The moments generated by this equation are determined to be 
\be
\langle z^{2n} \rangle =  \alpha \sum_{k=0}^{n-1} \frac{(2n-1)!}{2n + 1 - 2k} \frac{\langle z^{2k}\rangle}{ (2k)!}  \prod_{m=2k}^{2n-1} \frac{1}{m+\alpha}, 
\label{eq:zn2-3DL}
\ee
no longer a simple expression as that for two-dimensions.  Compare this expression with that in Eq. (\ref{eq:zn2d}).
To infer from those moments the first order differential equation that generates them is no longer straightforward and, 
at least, at this moment, we have no workable methodology to attain this goal.

We may try to get some insights by considering the limit $\alpha=0$, which is the case of quenched disorder.  
The distribution for a given drift orientation has Boltzmann form given in Eq. (\ref{eq:rho-alpha}).  
When integrated over all orientations, $\int_0^{\pi} d\theta\, \sin\theta \, \rho(x,\theta)$, in reduced units 
it becomes 
\be
p(z) \propto \text{erf} \left[\frac{z+1}{\sqrt{2 B}}\right]   -   \text{erf}\left[\frac{z-1}{\sqrt{2 B}}\right].
\ee
The third order differential equation that generates such a solution is found to be
\be
0   =     -    \big[1 - z^2 -   B\big] p'    +   2B z p''   +   B^2 p'''.   
\label{eq:diff-alpha-3DL}
\ee
Note the absence of the zero order term.  This should be of help when inferring a differential equation for $B=0$.

We return now to the case $B=0$ for which we propose the following ansatz differential equation:
\be
0  =   -\alpha  zp    -   \left (1- z^2\right) p'    -   \left[p\sum_{n=0}^M b_n z^{2n} \right]'.  
\ee
The coefficients $b_n$ and their number $M$ still must be determined.  Combining the ansatz with 
Eq. (\ref{eq:diff-alpha-3DL}) and adding the term $-\alpha B p'$ to ensure correct behavior in the limit 
$\alpha\to\infty$, the complete equation becomes 
\ba
0   &=&      -\left (\alpha + 2\sum_{n=0}^M b_n n z^{2n-2} \right) zp   \nonumber\\ 
&-&    \left(1 - z^2    -   B    +   \alpha B   +  \sum_{n=0}^M b_n z^{2n} \right ) p'    
+   2B z p''   +   B^2 p'''.   \nonumber\\
\label{eq:diff-all-3DL}
\ea
What we know about the coefficients $b_n$ is that in the limit $\alpha\to 0$ they vanish if we are to have an 
agreement with Eq. (\ref{eq:diff-alpha-3DL}).  We also know that in the limit $\alpha\to\infty$ they go to a finite value, 
to ensure convergence to passive Brownian particle behavior.

Finally, since for $B=0$ the effective diffusion constant, defined as 
$
D_{eff} =   1 -  z^2 + \sum_{n=0}^M b_n z^{2n}, 
$
should vanish at $z=1$ to prevent particles to exist beyond $z>0$, this suggests the constraint 
$$
\sum_{n=0}^M b_n = 0.  
$$
From Eq. (\ref{eq:diff-all-3DL}) in the limit $B=0$ we obtain the following relation for the moments:  
\be
0   =    \alpha \langle z^{2m} \rangle   -   (2m-1) \sum_{n=0}^M b_n \langle z^{2n+2m-2} \rangle.  
\label{eq:bn}
\ee
In combination with Eq. (\ref{eq:zn2-3DL}), in principle, we should be able to calculate the coefficients $b_n$.  

The hope is that $M$ is finite and the expressions for $b_n$ are simple.  This turns out not to be the case.  
An infinite number $b_n$ is needed if Eq. (\ref{eq:bn}) is to be satisfied for all $n$.  The expression for those 
coefficients are not simple and grow in complexity with increasing $M$.  

\section{summary and discussion}
\label{sec:sec5} 

Table (\ref{table1}) summarizes the main results of this work:  third-order equations for the flux $j$ 
for RTP particles in a harmonic trap.  
\begin{table}[h!]
\centering
 \begin{tabular}{l l } 
 \hline
 & \\[-1.8ex]
 &  ~~~~~~~~~~~~~~~~~~~~~~~~~~~~~~~~~~~ RTP  \\ [0.2ex] 
 \hline
 & \\[-1ex]
1D & {\scriptsize $ ~~  j =    \big[2-  \alpha\big] z p  -    \big[1 - z^2  -  3B  +   \alpha B\big] p'   +  2B  z  p''   +  B^2 p'''$}   \\ [1.ex] 
2DL &  {\scriptsize $ ~~ j =    \big[1-2\alpha\big] z p  -    \big[1 - z^2 -   2B  +  2\alpha B\big] p'    +   2B z p''   +   B^2 p'''$}  \\ [1.ex] 
2D &  {\scriptsize $ ~~ j =    \big[2-2\alpha\big] z p  -    \big[1 - z^2  -  4B  + 2\alpha B  +  \frac{B^2}{z^2} \big] p'  + \big[ 2 B  +  \frac{B^2}{z^2} \big] zp''  +  B^2 p''' $}  \\ [1.ex] 
  \hline
\end{tabular}
\caption{Third-order equations for flux $j$ of RTP particles in a harmonic trap. 
2DL denotes two-dimensional system with harmonic potential in $x$-direction, $u=Kx^2/2$, and 
2D denotes isometric harmonic potential, $u=Kr^2/2$.   For 1D and 2DL, $z=x/\lambda_k$ 
and for 2D, $z=r/\lambda_k$.  Apart for different parametrization, equation 1D and 2DL are the same.  
Equation 1D is transformed into equation 2DL using the transformation $\alpha\to  2\alpha + 1$.}
\label{table1}
\end{table}
The stationary distributions are obtained from the condition $j=0$ together with the boundary conditions
\ba
&& p'(0) = 0, \nonumber\\
&&  p(\pm \infty) = 0.  
\ea
The first condition ensures the symmetry of $p$ around $z=0$, and the second condition ensures that $p$ 
vanishes far from the trap center.  The normalization constraint of $p$ fully defines the solution.  
The normalization for 1D and 2DL is 
$
\int_{-\infty}^{\infty} dz\, p(z) = 1
$
and for 2D it is 
$
2\pi \int_{0}^{\infty} dz\, z p(z) = 1.
$

In the limit $\alpha\to\infty$, where the swimming orientation changes rapidly, all equations become dominated 
by the terms linear in $\alpha$ and reduce to $j = -z p - Bp'$, a system of passive Brownian particles in a harmonic 
trap.  

Third order equations in Table (\ref{table1}) do not lend themselves to a closed form solution, but the solutions can 
be represented as convolution of the beta and Gaussian distributions as listed in Table (\ref{table1a}).  
Such form of the solution is intuitively accurate and represents a sum of two independent random processes:
thermal fluctuations and active motion.  This separation, however, should not be regarded as universal.  It is rather 
a feature of a harmonic trap.  
\begin{table}[h!]
\centering
 \begin{tabular}{l l } 
 \hline
 & \\[-1.8ex]
 &  ~~~~~~~~~~~~~~~~~~~ RTP  \\ [0.2ex] 
 \hline
 & \\[-1ex]
1D & {\scriptsize $ ~~  p \propto \int_{-1}^{1} dz'\, (1-z'^2)^{\frac{\alpha}{2}-1} e^{-\frac{(z-z')^2}{2B}}.$}   \\ [1.ex] 
2DL &  {\scriptsize $ ~~ p \propto \int_{-1}^{1} dz'\, (1-z'^2)^{\alpha-\frac{1}{2}} e^{-\frac{(z-z')^2}{2B}}.$}  \\ [1.ex] 
2D &  {\scriptsize $ ~~ p \propto  \int_0^1  dz'\, (1-z'^2)^{\alpha-1} e^{-\frac{ (z' - z)^2}{2B}}  \left[ z'  e^{-zz'/B} {\text I}_0\left(\frac{z z'}{B}\right) \right] 
, $}  \\ [1.ex] 
  \hline
\end{tabular}
\caption{Solutions to third order equations in Table (\ref{table1}) for $j=0$.  }
\label{table1a}
\end{table}

The fact that all equations are third-order appears to be a hallmark of active motion, without counterpart
in passive Brownian dynamics.  Without thermal fluctuations, or $B=0$, all equations 
reduce to first-order for which a solution is readily available as indicated in Table (\ref{table2}).  
The onset of thermal fluctuations gives rise to second and third order terms.  
\begin{table}[h!]
\centering
 \begin{tabular}{l l l } 
 \hline
 & \\[-1.8ex]
 & ~~~~~~~~~~~~~~~~~~~ RTP, ~~$B=0$  \\ [0.2ex] 
\hline
 & \\[-1ex]
1D     &  {\scriptsize ~~~$j = (2-\alpha)zp - (1-z^2) p' $ }  &   {\scriptsize $ ~~~~~ p \propto (1-z^2)^{\frac{\alpha}{2} - 1} $ }    \\ [1.ex] 
2DL  &  {\scriptsize ~~~$j = (1-2\alpha)zp - (1-z^2) p' $}  &   {\scriptsize $ ~~~~~  p \propto (1-z^2)^{\alpha - \frac{1}{2}} $} \\ [1.ex] 
2D  &  {\scriptsize ~~~$j = (2-2\alpha)zp - (1-z^2) p' $}  &     {\scriptsize $ ~~~~~  p \propto (1-z^2)^{\alpha - 1} $}   \\ [1.ex] 
  \hline
\end{tabular}
\caption{As in Table (\ref{table1}) but for a zero temperature. }
\label{table2}
\end{table}


The solutions in Table (\ref{table2}) are beta distributions with different $\alpha$ scaling.  In each case 
the crossover $\alpha_c$, the point where the distribution changes from concave to convex, is different.  
The cases 1D and 2DL are "effectively" one-dimensional.  The difference between these two cases comes
from the distribution of swimming velocities.  For the 1D case, 
the swimming velocities are discrete, $v=\pm v_0$, and for the 2DL case they are continuously distributed 
in the interval $v\in [-v_0,v_0]$ (in this case only the $x$-projection of the swimming velocities is relevant).  
As a result, the swimming velocities at any time are smaller than $v_0$ and particles push less against the 
harmonic potential.  This should result in a less concave distribution for the same value of $\alpha$ compared
to the 1D case.  This is what is observed.  This also explains why crossover for 2DL occurs at $\alpha_c=1/2$
while that for 1D at $\alpha_c=2$.

We next compare the 1D and 2D cases.  First we note that for $\alpha=0$ the distributions are the same.  
This happens because for $\alpha=0$ swimming orientations do not change in time and so  
$\theta_r=\theta$, where $\theta_r$ is the angular position of a particle in the trap and $\theta$ is the 
swimming orientation. For finite $\alpha$, however, it is expected that 
$\theta_r \neq \theta$, this in turn results in the velocity component projected onto the 
direction $\theta_r$ to be less than $v_0$, or $v_0\cos(\theta-\theta_r) \le v_0$.


As the higher order terms do not vanish in the limit $\alpha=0$, see Table (\ref{table3}), 
these terms need not necessarily be attributed to some property of non-equilibrium.  
\begin{table}[h!]
\centering
 \begin{tabular}{l l } 
 \hline
 & \\[-1.8ex]
 &  ~~~~~~~~~~~~~~~~~~~~~~ RTP, ~~$\alpha=0$  \\ [0.2ex] 
 \hline
 & \\[-1ex]
1D & {\scriptsize $ ~~  j =    2z p  -    \big[1 - z^2  -  3B \big] p'   +  2B  z  p''   +  B^2 p'''$}   
\\ [1.ex] 
2DL &  {\scriptsize $ ~~ j =    z p  -    \big[1 - z^2 -   2B \big] p'    +   2B z p''   +   B^2 p'''$}    
\\ [1.ex] 
2D &  {\scriptsize $ ~~ j =    2z p  -    \big[1 - z^2  -  4B  +  \frac{B^2}{z^2} \big] p'  + \big[ 2 B z  +  \frac{B^2}{z} \big] p''  +  B^2 p''' $}    
\\ [1.ex] 
3DL &  {\scriptsize $ ~~ j =     -    \big[1 - z^2 -   B\big] p'    +   2B z p''   +   B^2 p'''$}    
\\ [1.ex] 
3D &  {\scriptsize $ ~~ j =    2 z p  -    \big[1 - z^2  -  5B  +  \frac{2B^2}{z^2} \big] p'  + \big[ 2 B z  +  \frac{2B^2}{z} \big] p''  +  B^2 p''' $}    
\\ [1.ex] 
  \hline
\end{tabular}
\caption{As in Table (\ref{table1}) but for $\alpha=0$.  This limit represents a system in equilibrium with 
quenched disorder.  Equations for three-dimensional case are included since they can be obtained using 
procedure in Appendix (\ref{sec:app3}).  }
\label{table3}
\end{table}
Since a system at $\alpha=0$
represents equilibrium with quenched disorder, higher order terms can be interpreted as arising from  
disorder due to different swimming orientations.  All equations can be written as a combination of two
equations representing different limits.  Taking the 1D case as an example, we have 
\ba
0  &=&    \left\{2z p  -    \big[1 - z^2  -  3B\big] p'   +  2B  z  p''  +  B^2 p'''  \right\} \nonumber\\
&+&   \alpha \left\{ -z p - Bp' \right\}. 
\ea
Both parts of the equation taken separately represent equilibrium:  equilibrium for quenched disorder 
and equilibrium of passive Brownian particles.  Parameter $\alpha$ determines the balance between the 
two equilibria.  

\section{Conclusion}
\label{sec:sec6}

In this work we posit the problem of stationary distributions of active particles in a harmonic trap as third-order 
linear homogenous differential equation.  Using the procedure developed in this work, we were able to determine 
such equation for RTP particles in 2D for different trap symmetries.  We did not obtain an analogous differential 
equations for RTP particles in higher dimensions.  Similar difficulties were encountered for the case of ABP 
particles in any dimension and for any symmetry of a trap.  In none of those cases it was possible to infer a 
differential equation for the limit $B=0$ from even moments $\langle z^2 \rangle$.  More fundamental causes 
of why a simple differential equation is possible for RTP particles in 2D and not for other cases remains to be 
better understood.  

In addition to formulating the problem as third order equation, we obtain solutions in the form of convolution 
of two probability distributions.  This is reflection of the fact that the two random processes, thermal fluctuations 
and active motion, are independent and the total distribution is for the variable that is a sum of those two processes.  
Retrospectively, such explanation makes perfect sense.  On the other hand, we should not expect that it 
applies to other types of external potentials.  It is rather a characteristic of a harmonic potential.

\begin{acknowledgments}
I am indebted to Jean-Marc Luck for recognizing that the solution to third order differential equation 
for the 1D case can be 
represented as a convolution of beta and Gaussian distributions.  
D.F. acknowledges financial support from FONDECYT through grant number 1201192.  
\end{acknowledgments}

\section{DATA AVAILABILITY}
The data that support the findings of this study are available from the corresponding author upon 
reasonable request.

\appendix

\section{Integral solutions to third order differential equations}
\label{sec:app0}

In this section we derive a solution to Eq. (\ref{eq:FP0-harmonic}) given in the form of convolution, introduced 
in the main text in Eq. (\ref{eq:p1D-conv}).  For convenience, we provide Eq. (\ref{eq:FP0-harmonic}) below:  
$$
0 =    (2-\alpha)zp    -   (1-z^2  - 3B  +  B\alpha)p'    +    2B zp''    +   B^2 p'''. 
$$
The above equation is transformed by the application of two-sided Laplace transform \cite{JML22}:  
\be
\hat p = \int_{-\infty}^{\infty} dz\, e^{-sz} p, 
\label{eq:app0a}
\ee
leading to 
\be
0 =     (1  +  B  +  B\alpha - B^2 s^2  ) s \hat p   +  (2B s^2 - \alpha)\hat p'      -   s \hat p''.
\label{eq:app0b}
\ee
If the transformed solution can be represented as $\hat p(s) = K_{\alpha}(s) K_B(s)$, where $K_{\alpha}(s)$ depends 
only on the parameter $\alpha$ and $K_B(s)$ only on the parameter $B$, then 
\be
K_{\alpha}(s) K_{B}(s) = \int_{-\infty}^{\infty} dz\, e^{-sz} p(z), 
\ee
which implies that $p$ is the convolution of the inverse transforms of $K_{\alpha}(s)$ and $K_B(s)$: 
\be
p(z) \propto \int_{-1}^{1} dz'\, K^{-1}_{\alpha}(z') K^{-1}_B(z-z').  
\label{eq:app0c}
\ee
Our next task is to determine what the functions $K_{\alpha}$ and $K_B$ are.  

To find $K_{\alpha}$, we set $B=0$ so that Eq. (\ref{eq:app0b}) reduces to 
\be
0 =     s \hat p   - \alpha\hat p'      -   s \hat p''.  
\label{eq:app0d}
\ee
But since we know that the distribution $p(z)$ in this case is $p = (1-z^2)^{\frac{\alpha}{2}-1}$, the solution to 
Eq. (\ref{eq:app0d}) must be 
\be
\hat p \propto \int_{-1}^{1} dz\, e^{-sz} (1-z^2)^{\frac{\alpha}{2}-1}.  
\label{eq:app0e}
\ee
We are next going to claim that for $B=0$, $K_B=1$  (or some other constant) so that
\be
K_{\alpha}(s)  \propto \int_{-1}^{1} dz\, e^{-sz} (1-z^2)^{\frac{\alpha}{2}-1}.  
\label{eq:app0f}
\ee
We will be able to verify this assumption later when we have an expression for $K_B$.  

To determine $K_B$, we want to identify the limit where $K_{\alpha}$ becomes independent of $s$, namely 
$K_{\alpha}=1$ (or some other constant), so that $\hat p=K_B$.  Examining Eq. (\ref{eq:app0f}) we identify 
this to be $\alpha\to \infty$, where $(1-z^2)^{\frac{\alpha}{2}-1}$ becomes a delta function and $K_{\alpha}$ 
reduces to a constant.  Eq. (\ref{eq:app0b}) for this limiting situation reduces to 
\be
0 =     B s \hat p   - \hat p',
\label{eq:app0g}
\ee
where the solution is $\hat p = e^{B s^2/2}$.  Based on this result, we get 
\be
K_{B}(s) \propto e^{\frac{Bs^2}{2}} \propto  \int_{-\infty}^{\infty} dz\, e^{-sz}  e^{-z^2/2 B}.  
\ee
Going back to Eq. (\ref{eq:app0c}), we can write 
\be
p(z) \propto \int_{-1}^{1} dz'\, (1-z'^2)^{\frac{\alpha}{2}-1} e^{-(z'-z)^2/2 B}.  
\label{eq:app0h}
\ee
This is the solution first given in Eq. (\ref{eq:p1D-conv}).

To obtain the solution in Eq. (\ref{eq:p2D-conv}) for the harmonic potential in 2D, we assume that the 
solution has a similar form to that in Eq. (\ref{eq:app0e}), except the convolution is over 
the 2D space:
\be
p \propto e^{-\frac{z^2}{2B}}  \int  d{\bf z}'\, (1-z'^2)^{\alpha-1} e^{ -\frac { ({\bf z}' - {\bf z})^2} {2B} }, 
\ee
where ${\bf z}$ and ${\bf z}'$ are the vectors in 2D.  
Using polar coordinates, this becomes 
$$
p \propto e^{-\frac{z^2}{2B}}  \int_0^1  dz'\, z' (1-z'^2)^{\alpha-1} e^{-\frac{ z'^2}{2B}}  \int_0^{2\pi} d\theta\, e^{\frac{zz'\cos\theta}{B}},
$$
which after integration over $\theta$ and some rearrangement becomes 
\be
p \propto  \int_0^1  dz'\, (1-z'^2)^{\alpha-1} e^{-\frac{ (z' - z)^2}{2B}}  \left[ z'  e^{-zz'/B} {\text I}_0\left(\frac{z z'}{B}\right) \right] .  
\ee
This can be verified to be a solution to Eq. (\ref{eq:p_2DS}).

\section{Procedure for reducing Eq. (\ref{eq:FPT-free-2})}
\label{sec:app1}

This section presents the procedure for reducing two coupled equations in (\ref{eq:FPT-free-2}) 
into Eq. (\ref{eq:FPT-free-1}) for the distribution $p=p_+ + p_-$.   We begin by adding and subtracting 
the two equation in Eq. (\ref{eq:FPT-free-2}).  This procedure transforms the two equations into 
\ba
&& \dot p  =  -v_0 \sigma'   +   D p''  , \nonumber\\
&& \dot \sigma   =   -v_0 p'   +   D \sigma''   -    \frac{\sigma}{\tau},
\label{eq:app1a}
\ea
where $\sigma=p_+ - p_-$.  The first equation in (\ref{eq:app1a}) is used to obtain a number of expressions 
for derivatives of $\sigma$ in terms of $p$ and its derivatives:   
\ba
&&  v_0 \sigma'    =    D p''   -    \dot p      \nonumber\\
&&  v_0 \sigma'''    =    D p''''   -    \dot p''      \nonumber\\
&&  v_0 \dot \sigma'    =    D \dot p''   -    \ddot p.    
\label{eq:app1b}
\ea
By differentiating the second equation in (\ref{eq:app1a}) with respect to $x$ we get 
\be
\dot \sigma'   =   -v_0 p''   +   D \sigma'''   -    \frac{\sigma'}{\tau}.  
\ee
The expressions in (\ref{eq:app1b}) are then used to eliminate all terms with $\sigma$:   
\be
\dot p  -   2\tau D \dot p''   +    \tau \ddot p    =    (D + \tau v_0^2) p''    -   \tau D^2 p''''.  
\ee
The result agrees with Eq. (\ref{eq:FPT-free-1}) and Eq. (\ref{eq:flux-free}).

\section{Procedure for reducing Eq. (\ref{eq:FP_RTP1D}) }
\label{sec:app2}

This section presents the procedure for reducing the two coupled equations in (\ref{eq:FP_RTP1D}) 
into a single equation in (\ref{eq:FPT-harmonic-1}).  We start by transforming those equations into 
\ba
&& \dot p  =   zp'   - \sigma' + B p''  +  p, \nonumber\\
&& \dot \sigma  =   z\sigma'  - p'   + B\sigma''   +   (1-\alpha) \sigma.
\label{eq:app2a}
\ea
where $p=p_+ + p_-$ and $\sigma=p_+ - p_-$.  
The first equation in (\ref{eq:app2a}) is used to obtain a number of expressions 
for different derivatives of $\sigma$ in terms of $p$ and its derivatives:  
\ba
&& \sigma'   =   zp'    +   B p''   +   p   -  \dot p        \nonumber\\
&& \sigma''   =   zp''    +   B p'''   +   2p'   -  \dot p'  \nonumber\\
&& \sigma'''   =   p''    +    zp'''    +   B p''''   +   2p''   -  \dot p''  \nonumber\\
&& \dot \sigma'   =   z\dot p'    +   B \dot p''   +   \dot p   -  \ddot p.  
\label{eq:harmonic-formulas}
\ea
Next, we differentiate the second equation in (\ref{eq:app2a}) with respect to $z$:
\be
 \dot \sigma'  =  \sigma'  +  z\sigma''  - p''   + B\sigma'''   +   (1-\alpha) \sigma'.  
\ee 
The terms involving $\sigma$ are next eliminated using the expressions in (\ref{eq:harmonic-formulas}): 
\ba
 (\alpha-3) \dot p  -  2z\dot p'    -   2B \dot p''     +  \ddot p  &=&     (\alpha-2) p  -  (4-\alpha) zp'   \nonumber\\
&-&  (z^2  - 1 +  5B - B\alpha) p''   \nonumber\\
& -&   2B z p'''    -  B^2 p''''.    \nonumber
 \label{eq:app2b}
 \ea
To convert the rhs of the equation to conform with the expression $-j'$, we rewrite it as 
\ba
rhs &=& -(2-\alpha) p  -  (2-\alpha) zp'   - 2zp'   \nonumber\\ 
&-&   (- 1 +  3B - B\alpha) p''  -  z^2p'' -  2Bp''  \nonumber\\ 
 &-&   2B z p'''    -  B^2 p''''.
\ea
This can be rearranged into
\be
rhs = -\left[ (2-\alpha) zp    +  (z^2 - 1 +  3B - B\alpha) p' +  2Bzp''   +  B^2 p''' \right]'.  
\label{eq:app2c}
\ee
Eq. (\ref{eq:app2b}) and Eq. (\ref{eq:app2c}) yield the results in 
Eq. (\ref{eq:FPT-harmonic-1}) and Eq. (\ref{eq:flux-harmonic}).

\section{Fokker-Planck equations for RTP particles in a harmonic trap:  derivation of Eq. (\ref{eq:FP0-2DL}) and Eq. (\ref{eq:FPS-2Da})}
\label{sec:app4}

The FP equation for the RTP model in 2D subject to an external force ${\bf F}$ is 
\be
\dot\rho =  -\bnabla \cdot \left[ \left(\mu {\bf F} + v_0  {\bf n} \right) \rho \right]  
+ D \nabla^2 \rho 
-  \frac{1}{\tau} \left( \rho - \int_0^{2\pi} \frac{d\theta}{2\pi}\, \rho \right).  
\ee
For ${\bf F} = - {\bf e}_x Kx$ and at steady-state $\rho\equiv \rho(x,\theta)$, and the FP equation becomes 
\be
0 =  \frac{\partial}{\partial x} \left[ \left(\mu K x - v_0  \cos\theta \right) \rho \right]
+ D \frac{\partial^2\rho}{\partial x^2} -  \frac{1}{\tau} \left( \rho - \int_0^{2\pi} \frac{d\theta}{2\pi}\, \rho \right).  
\ee
Using dimensionless units, this recovers Eq. (\ref{eq:FP0-2DL}).

In the case of an isotropic force ${\bf F} = -K{\bf r}$, at steady-state the FP equation becomes 
\ba
0 &=&  \mu K  {\bf r}  \cdot \bnabla \rho  +  \mu K \rho (\bnabla \cdot {\bf r}) - v_0  \bnabla \cdot  {\bf n} \rho \nonumber\\ 
&+& D \nabla^2 \rho  -  \frac{1}{\tau} \left( \rho - \int_0^{2\pi} \frac{d\theta}{2\pi}\, \rho \right).  
\ea
The stationary $\rho$ should be a function of the position of a particle in a trap, in polar coordinates
this includes $r$ and $\theta_r$, and the swimming orientation $\theta$.   Actually, the only relevant angular 
dependence is the difference $\theta_r-\theta$.  This allows us to represent a stationary $\rho$ as a function of
two variables.  From now on we use $\theta$ to designate the difference $\theta_r-\theta$.  

For convenience, we fix the swimming orientation ${\bf n}$ along the $x$-axis, ${\bf n} = {\bf e}_x$. 
The resulting equation becomes 
\ba
0 &=&  \mu K  {\bf r}  \cdot \bnabla \rho  +  \mu K \rho (\bnabla \cdot {\bf r}) - v_0  \frac{\partial \rho}{\partial x}  \nonumber\\ 
&+& D \nabla^2 \rho  -  \frac{1}{\tau} \left( \rho - \int_0^{2\pi} \frac{d\theta}{2\pi}\, \rho \right).  
\ea

In polar coordinates various terms are expressed as 
$\nabla^2\rho = \frac{\partial^2 \rho}{\partial r^2} + \frac{1}{r}  \frac{\partial \rho}{\partial r} + \frac{1}{r^2} \frac{\partial^2 \rho}{\partial\theta^2}$,
$\frac{\partial \rho}{\partial x} = \cos\theta \frac{\partial\rho}{\partial r} -   \frac{\sin\theta}{r} \frac{\partial \rho}{\partial \theta}$,
${\bf r} \cdot \bnabla \rho  = r \frac{\partial \rho}{\partial r}$, $\bnabla \cdot {\bf r} = 2$.  Substituting those expressions in the 
above equation leads to 
\ba
0 &=&  \mu K r \frac{\partial \rho}{\partial r}  +  2 \mu K \rho  - v_0  \cos\theta \frac{\partial\rho}{\partial r} + v_0  \frac{\sin\theta}{r} \frac{\partial \rho}{\partial \theta}   \nonumber\\ 
&+& D \left[  \frac{\partial^2 \rho}{\partial r^2}       +   \frac{1}{r}  \frac{\partial \rho}{\partial r}     +   \frac{1}{r^2} \frac{\partial^2 \rho}{\partial\theta^2}  \right]  
-  \frac{1}{\tau} \left( \rho - \int_0^{2\pi} \frac{d\theta}{2\pi}\, \rho \right).  \nonumber\\
\ea
Using dimensionless units, this recovers Eq. (\ref{eq:FPS-2Da}).

\section{Reverse procedure of inferring equation from solution}
\label{sec:app3}

In this section we outline a procedure for inferring equation from solution.  The
procedure is used to obtain Eq. (\ref{eq:diff-2DL-q}) and Eq. (\ref{eq:p_2DS-B}) from the solution 
\be
p(z) \propto  e^{ - \frac{z^2}{2B} }  \int_{0}^{2\pi} d\theta\,   e^{-\frac{ \cos^2\theta}{2 B}}  e^{  \frac{z \cos\theta }{B} } 
\label{eq:app3a}
\ee
first appears in Eq. (\ref{eq:p2DL-q}), and 
\be
p(z) \propto e^{-z^2/2B} \,\,  {\text I}_0(z/B), 
\label{eq:app3aa}
\ee
that appears in Eq. (\ref{eq:p0}).   As both functions are even, their Taylor expansion 
$$
p(z) = p^{(0)}(0) + \frac{z^2}{2} p^{(2)}(0)  + \frac{z^4}{4!} p^{(4)}(0)  + \dots.  
$$
We next propose the following ansatz for an unknown third-order equation: 
\ba
0 &=&   (c_{11} + c_{21} z^2 + c_{31} z^4) zp     \nonumber\\
   &+&   (c_{12} + c_{22} z^2 + c_{32} z^4) p'    \nonumber\\
   &+&   (c_{13} + c_{23} z^2 + c_{33} z^4) zp'' \nonumber\\ 
   &+&   (c_{14} + c_{24} z^2 + c_{34} z^4) p'''.  
\label{eq:app3b}
\ea
Note that ansatz does not introduce mixed even and odd terms.  This is to ensure that symmetry of $p$ is preserved.  

The coefficients $c_{ij}$ are determined from a truncated expansion of $p$, for example
$$
p_{tr} = p^{(0)} + \frac{z^2}{2} p^{(2)} + \frac{z^4}{4!} p^{(4)} +  \frac{z^6}{6!} p^{(6)},
$$
where $p^{(0)} =  p(0)$, $p^{(1)} =  p'(0)$, etc.  The number of terms in $p_{tr}$ depends on the 
number of parameters $c_{ij}$.  For example, for nine $c_{ij}$, we should include 
at least nine terms in the truncated series (actually, eight is enough since we assume $c_{11}=1$).  

If the coefficients do not become alterened by using $p_{tr}$ for larger number of terms, it is assumed that the equation 
is complete.    For the distribution in (\ref{eq:app3a}), the nonzero coefficients are 
\ba
&& c_{11} =  1,     \nonumber\\
&& c_{12} =  2B - 1, ~~ c_{22} = 1,    \nonumber\\
&& c_{13} =  2 B,   \nonumber\\
&& c_{14}=  B^2.  
\label{eq:app3c}
\ea
This suggests the following equation 
\be
0 = z p   -   (1 - z^2 - 2B) p'    +   2 B z p''   +   B^2 p''', 
\label{eq:app3d}
\ee
which can be verify to yield the solution in Eq. (\ref{eq:app3a}).  

For the distribution in (\ref{eq:app3aa}), the nonzero coefficients are 
\ba
&& c_{21} = 2,  \nonumber\\
&& c_{12} =  -B^2, ~~ c_{22} = 4B - 1, ~~ c_{32} = 1,     \nonumber\\
&& c_{13} =  B^2, ~~  c_{23} = 2B,   \nonumber\\
&& c_{24} = B^2, 
\label{eq:app3c}
\ea
and after manipulation, the ansatz becomes  
\be
0 =   2 z p  + \bigg( z^2 -1 +  4B  -  \frac{B^2}{z^2}\bigg) p'  +   \bigg(\frac{B^2}{z^2} + 2B \bigg) z p''  +  B^2 p'''.
\ee







\end{document}